\documentstyle[aps]{revtex}
\begin{document}
\draft 
\preprint{} 
\title{Dynamical Phase Transition in a Fully Frustrated Square Josephson Array} 
\author{K. D. Fisher, D. Stroud, and L. Janin}
\address{Department of Physics, The Ohio State University, Columbus,
Ohio 43210}
\date{\today}
\maketitle

\begin{abstract}
We study dynamical phase transitions at temperature $T = 0$ in a fully
frustrated square Josephson junction array subject to a driving
current density which has nonzero components $i_x$, $i_y$ parallel to
{\em both} axes of the lattice.  Our numerical results show clear
evidence for {\em three} dynamical phases: a pinned vortex lattice
characterized by zero time-averaged voltages $\langle v_x\rangle_t$
and $\langle v_y \rangle_t$, a ``plastic'' phase in which both
$\langle v_x\rangle_t$ and $\langle v_y\rangle_t$ are nonzero, and a
moving lattice phase in which only one of the time-average voltage
components is nonzero.  The last of these has a finite transverse
critical current: if a current is applied in the $x$ direction, a
nonzero transverse current density $i_y$ is required before $\langle
v_y\rangle_t$ becomes nonzero.  The voltage traces in the moving
lattice phase are periodic in time.  By contrast, the voltages in the
plastic phase have continuous power spectra which are weakly dependent
on frequency.  This phase diagram is found numerically to be
qualitatively unchanged by the presence of weak disorder.  We also
describe two simple analytical models which recover some, but not all,
the characteristics of the three dynamical phases, and of the phase
diagram calculated numerically.

\end{abstract}
\pacs{74.60.Ge,74.76.-w,74.50.+r,74.60.Jg}

\section{Introduction}

\label{sec:intro}

There has recently been considerable interest in dynamical phases in
superconductors.  Such interest has been stimulated by much evidence
(both theoretical and experimental) that vortex lattices, as well as
other ordered systems, can exhibit various types of transitions from
one type of dynamical phase to another, as a function of controllable
external parameters, such as driving force and temperature.  In the
case of Type-II superconductors in a magnetic field, the driving force
which acts on the vortex lattice is the applied current.

In disordered superconducting films and crystals, there now appear to
exist at least three distinct phases, as a function of driving current
and temperature\cite{bhatta,yaron,marchevsky,heller,henderson,dilley}.
At low driving current, the vortex lattice is pinned, and typically
exhibits a glass-like order because of the random pinning centers
which prevent the lattice from moving.  At intermediate driving
currents, the glass phase is depinned, and starts to move; this motion
is thought to occur inhomogeneously in disordered superconductors
(that is, some of the vortices move through random channels, while
other vortices remain pinned).  In this phase, the vortex system is
usually said to exhibit {\em plastic flow}.  Finally, at high driving
currents, an event resembling a phase transition occurs, and the
vortex system reverts back to an ordered state which closely resembles
a moving crystalline phase.  Most workers now believe that this third
phase, especially in two dimensions, lacks long-range crystalline
order but instead exhibits hexatic or smectic order, modified by the
static disorder of the system\cite{balents,spencer}.  This
high-current phase is also thought to exhibit a {\em finite transverse
critical current}, i.\ e., once moving parallel to the driving force,
the lattice requires a finite driving force perpendicular to the
average direction of motion before acquiring a nonzero transverse
velocity.  The details of all these phases, as well as the transitions
between them, remain the subjects of much experimental and theoretical
investigation.

Recently, there has been evidence that some similar behavior is to be
found in {\em ordered} systems, both in two and three dimensions (2D
and 3D).  Extensive numerical studies have been carried out by Nori
and collaborators, who have found a very complex structure of ordered
and disordered (sometimes called ``plastic'') phases as a function of
the relative density of vortices and pinning sites, and of the
direction of applied current relative to the axes of the periodic
pinning lattice\cite{nori,nori1}.  Some of the observed structure can
be described in terms of nomenclature frequently used for
incommensurate dynamical systems, such as devil's staircases, Ar'nold
tongues, etc.\ These workers use a standard model whereby the vortices
interact with each other via a pairwise potential involving a modified
Bessel function, and with the pins by a suitable short-range pinning
potential.  A similar model had been used previously to study vortex
lattices driven through a {\em disordered} array of
pins\cite{ryu96}. This earlier calculation revealed a somewhat simpler
structure with three phases apparent: a pinned vortex lattice, a
moving plastic phase, and a moving vortex lattice with a finite
transverse critical current.

In this paper, we investigate the possibility of dynamical phase
transitions in a widely investigated model system: the overdamped
square Josephson junction array in a transverse magnetic field.  This
study is complementary to the work mentioned above, in that the phase
angles of the complex superconducting order parameter are explicitly
included as degrees of freedom.  Although such ordered Josephson
junction arrays have, of course, been extensively studied\cite{jja},
little attention has been paid to the possibility of plastic phases in
these materials.  Strongly disordered arrays have recently been shown
numerically to have both plastic and moving-lattice phases separated
by a phase boundary which appears to exhibit critical phenomena
associated with a diverging correlation length\cite{dominguez}.
Weakly disordered driven periodic systems are also predicted to have a
nearly periodic temporal order at large values of the driving
parameter\cite{balents1}.

In order to search for possible dynamical phases, we study this system
in a regime in which a current is applied with nonzero components
parallel to {\em both} the $x$ and $y$ axes of the array.  We consider
only the so-called fully-frustrated case with exactly 1/2 flux quantum
per plaquette of the square array.  Even in this ordered array, we
find three phases.  These are ($A$) the pinned vortex lattice, ($B$) a
moving lattice, and ($C$) a plastic phase occurring at currents
between $A$ and $B$.  Phases $A$ and $B$ have been much discussed
previously, but phase $C$ seems not to have been observed in an
ordered lattice.  The plastic phase $C$ is characterized by aperiodic
voltage noise and an absence of long range vortex order.  By contrast,
the moving lattice $B$ is characterized by voltage signals which are
periodic in time, and, most strikingly, by a finite transverse
critical current.

In an attempt to understand our numerical results, we also present
some simple analytical models.  One model is based on the assumption
that the dynamics can be characterized in terms of periodically
repeating 2 $\times$ 2 unit cells driven by a current applied at an
angle to the cell axes.  A second model considers the motion of a
single vortex through the periodic potential formed by the lattice of
Josephson-coupled grains.  Both models give rise to distinct regimes
in which none, one, or both of the voltage components parallel to the
cell axes are nonzero.  Thus, in particular, a nonzero transverse
critical current is present in both analytical models.

The remainder of this paper is organized as follows.  In Section II,
we briefly review our calculational model.  Our numerical results are
presented in Section III.  Section IV is an interpretive discussion
which includes the two analytical models mentioned above.  Our
conclusions are described in Section V.

\section{Model}

\label{sec:model}

In all our simulations, we take as the starting point the well-known
dynamical equations for an overdamped Josephson junction array in a
transverse magnetic field.  We assume that each Josephson junction is
resistively shunted, and we neglect the loop inductance; that is, we
assume that the magnetic field generated by the currents is negligible
compared to the applied magnetic field.  The equations then take the
form
\begin{equation}
I_{ab} = I_{c,ab} \sin{\left(\phi_a-\phi_b-A_{ab}\right)} +
        \frac{\hbar}{2eR_{ab}}\frac{d}{dt}\left(\phi_a-\phi_b-A_{ab}\right),
\end{equation}
\begin{equation}
\sum_b{I_{ab}} = I_a^{ext}.
\end{equation}
Here $I_{c,ab}$ is the critical current of the junction connecting
grains $a$ and $b$, $R_{ab}$ is the shunt resistance of that junction,
and $I_a^{ext}$ is the external current fed into the $a$th grain. We
assume a constant, uniform external field $\textbf{B} = B\hat{z}$
perpendicular to the array, and make the gauge choice
$\textbf{A}=Bx\hat{y}$.  The phase factor
\begin{equation}
A_{ab} = \frac{2\pi}{\Phi_0}\int{\textbf{A}\cdot d\textbf{s}}
\end{equation}
is then easily expressed in terms of $f=Ba^2/\Phi_0$, the frustration
or flux per plaquette of dimension $a \times a$, where $\Phi_0 = hc/2e$
is the flux quantum.

The equations of motion can be put in dimensionless form with the
definitions $i_{ab} \equiv I_{ab}/I_c$, $i_{c;ab} \equiv
I_{c,ab}/I_c$, and $g_{ab} \equiv R/R_{ab}$, and the natural time unit
$\tau \equiv \frac{\hbar}{2eRI_c}$, where $I_c$ and $R$ are a typical
critical current and a typical shunt resistance.  The result of this
substitution is
\begin{equation}
i_{ab} = i_{c,ab} \sin{\left(\phi_a-\phi_b-A_{ab}\right)} + \tau
        g_{ab}\frac{d}{dt}\left(\phi_a-\phi_b\right),
\end{equation}
\begin{equation}
\sum_b{i_{ab}} = i_a^{ext}.
\end{equation}
Combining these equations yields a set of coupled differential
equations which is easily reduced to matrix form and solved
numerically\cite{yu}.  In our work, we employed a fourth-order
embedded Runge-Kutta integration with variable time step\cite{karp}.

In most of our calculations, we have considered a lattice without
disorder: all shunt resistances $R_{ab} = R$ (or $g_{ab} = 1$), and
all critical currents $i_{c,ab} = 1$ on a square array.  Our array is
driven by nonzero current densities in both the $x$ and $y$
directions: a current $i_x$ is fed into each grain along the
left-hand edge of the array, and extracted from each grain on the
right-hand edge, and a current $i_y$ is similarly injected into each
grain on the bottom edge and extracted from each grain on the top edge
of the array (see Fig.\ \ref{fig:geom}).

Starting from randomized initial phases, we integrate these equations
of motion over an ``equilibration'' interval of 100$\tau$-5000$\tau$,
followed by an averaging period of 100$\tau$-2000$\tau$. Typically,
the external currents $i$ are ramped up or down (in steps of 0.001 to
0.1) without rerandomizing the phases.  We calculate the spatially
averaged but time-dependent voltage difference $v(t^\prime)=
V(t^\prime)/NRI_c$ (where $t^\prime=t/\tau$ is the dimensionless time)
between the input and output edges of the array in both the $x$ and
$y$ directions, as well as its time-average $\langle v\rangle_t$, in
both directions.  In some regions of the phase diagram, these voltages
appear to be periodic in time, as revealed by an analysis of the power
spectrum of the voltage.  In other regions, as described below, this
power spectrum reveals that the voltage is aperiodic.  In some of our
simulations, we also tracked the number and motion of vortices in the
array.  The vortex number in a given plaquette $\alpha$ is an integer
$n_{v,\alpha}$ defined by the relation
\begin{equation}
n_{v,\alpha} \equiv \frac{1}{2\pi} \sum_{ab}{(\phi_a-\phi_b-A_{ab})} =
0,\pm1,
\end{equation} 
where the sum is taken clockwise around the $\alpha^{th}$ plaquette,
and each phase difference is restricted to the range [$-\pi,\pi$].
Our calculations are carried out exclusively for a frustration $f =
1/2$, i.\ e., an applied magnetic field equal to one flux quantum for
every two plaquettes.  In a square Josephson array, the ground state
of this field is the well-known checkerboard vortex pattern, shown
schematically in Fig.\ \ref{fig:geom}.  Our simulations reproduce this
pattern.

\section{Results}

\label{sec:results}

The central results of our calculations are summed up concisely in
Fig.\ \ref{fig:phased}, which shows the ``dynamical phase diagram''
for two square Josephson junction arrays with no disorder at $f =
1/2$, driven by two orthogonal currents $i_x$ and $i_y$.  We find {\em
three different phases}: a pinned vortex lattice (time-averaged
voltages $v_x = v_y = 0$), a plastic flow regime ($v_x = 0$, $v_y > 0$
or $v_y = 0$, $v_x > 0$), and a moving vortex lattice ($v_x > 0$, $v_y
> 0$).  The calculated boundaries are shown in Fig.\ \ref{fig:phased}.

The results of Fig.\ \ref{fig:phased} were obtained by two different
methods. In the first procedure, the longitudinal current $i_x$ was
first ramped up from zero to a finite value, with $i_y$ held at zero.
Next, the transverse current $i_y$ was ramped up at fixed $i_x$.  To
determine the phase boundaries, we simply searched for the currents at
which the time averages $\langle v_x\rangle_t$ or $\langle
v_y\rangle_t$ (or both) became nonzero.  In the second method, we
ramped up $i_x$ and $i_y$ simultaneously, holding the ratio $i_y/i_x$
fixed.  Both methods generally gave similar phase
boundaries. Likewise, we found little indication of substantial
hysteresis, except in determining the boundary between phases $A$ and
$C$.  In this case, if the integration time is too long, the system
tended to jump abruptly back and forth between the two phases.

We refer to the phase in which both voltages are nonzero as a
``plastic'' phase, by analogy with a similar phase exhibited by
vortices in systems with quenched disorder.  In this phase,
$v(t^\prime)$ is apparently non-periodic in time; the corresponding
voltage power spectrum is only weakly frequency-dependent (see below).
Although we use the nomenclature of plastic, we have not checked that
the vortex motion in this phase is inhomogeneous (i.\ e.\ that only
some vortices are in motion while others remain pinned in this phase).
Such inhomogeneous motion is thought to occur in disordered
systems\cite{ryu96}.  By contrast, in the driven lattice phase, where
only $\langle v_x\rangle_t$ or $\langle v_y\rangle_t$ is nonzero, the
power spectra of the voltage traces are sharply peaked at a
fundamental frequency and its harmonics. Further, while one voltage is
always nonzero, the other voltage is periodic, only {\it averaging} to
zero over a cycle. We interpret this behavior as representing a vortex
lattice being driven transverse to the larger of the two current
components.

Various minor numerical difficulties sometimes interfered with the
calculations, but they could usually be overcome.  For example,
spurious voltage jumps were occasionally observed during these
calculations; these jumps (unlike genuine jumps) could generally be
eliminated by changing the initial conditions, the integration time,
or the direction of current ramping.  Only those voltage jumps which
appeared at the same position on the phase diagram in different runs
were deemed to be genuine.  From the occurrence of these jumps,
however, we conclude that at certain points on the phase diagram there
are a several metastable dynamical states which have similar energies.
The occurrence of such states may suggest a first-order transition
across the phase boundary, at least in the ordered system.

In Fig.\ \ref{fig:Vtrace}, we show time-dependent voltage traces at
several points in the plastic and moving-lattice phases. The traces
are plainly very different in the two phases.  In the moving-lattice
phase, the voltage traces are evidently periodic in time.  By
contrast, in the plastic phase, the voltages, both in $x$ and $y$
directions, are obviously aperiodic.  Another striking feature is
apparent in the moving lattice phase.  In this phase, as noted above,
there is a non-zero time averaged voltage only along one of the two
directions, even though current is applied along both the $x$ and $y$
directions.  Despite this feature, there is a finite {\em
time-dependent} voltage in the $y$ direction, which averages to zero,
and which is periodic like $v_x(t^\prime)$.

To learn more about the harmonic content of these voltage traces, we
have also calculated the {\em voltage power spectrum} $P(\omega\tau)$,
using the non-normalized Lomb method for variable-time-step
data\cite{recipes}:
\begin{equation}
P(\omega\tau) = 1/2 \left\{
\frac{\left[\sum_j\left(v_j-\bar{v}\right)\cos{\omega\left(t_j-t_o\right)}\right]^2}
{\sum_j\cos^2{\omega\left(t_j-t_o\right)}} +
\frac{\left[\sum_j\left(v_j-\bar{v}\right)\sin{\omega\left(t_j-t_o\right)}\right]^2}
{\sum_j\sin^2{\omega\left(t_j-t_o\right)}}\right\},
\end{equation}
where $t_o$ is defined by
\begin{equation}
\tan{\left(2\omega t_o\right)} = \frac{\sum_j{\sin{2\omega
t_j}}}{\sum_j{\cos{2\omega t_j}}}.
\end{equation}
Here, $\omega$ is the angular frequency, the $t_j$'s are the times at
which the voltage is recorded, $v_j=v\left(t_j\right)$, and $\bar{v}$
is the arithmetic average of the $v_j$'s. We have carried out this
calculation for several points in the plastic and moving-lattice
phases as indicated by triangles in Fig.\ 2(b).  The results are shown
in Fig.\ \ref{fig:Vpow} for both $v_x$ and $v_y$.  Clearly, the noise
confirms that the voltage in the moving-lattice phase is periodic in
time, while that in the plastic phase has a continuous spectrum which
is relatively weakly dependent on frequency.

The moving-lattice phase is characterized by a fundamental angular
frequency $\omega_0$.  It is readily shown that $\omega_0$ is related
to the time-averaged voltage drop across the array, $\langle V
\rangle_t$, by
\begin{equation}
\omega_0 = 2e\langle V \rangle_t/\hbar
\end{equation}
This relation is consistent with the widely-accepted egg-carton
picture of the vortex lattice in the moving phase\cite{lobb}.  In this
picture, the vortex lattice is viewed as a collection of eggs moving
in a potential similar to an egg-carton, consisting of a periodic
distribution of wells on a square lattice (each well lying at the
center of a plaquette formed by four grains).  During one period, the
vortex lattice moves by one row through the egg-carton potential.
Since there is a phase slip of $2\pi$ between two opposite edges of
the array each time a vortex crosses the line joining those two edges,
one can readily deduce the above relationship.

Fig.\ \ref{fig:phased}(a) shows that a finger of the driven lattice
phase is interposed between the pinned and plastic flow phases in a
$10 \times 10$ array.  This finger appears to be a finite-size
artifact, because it is absent in the phase diagram for a $20 \times
20$ array shown in Fig.\ \ref{fig:phased}(b).  We believe that the
phase diagram of Fig.\ \ref{fig:phased}(b) is likely to persist in an
$N \times N$ lattice even at very large $N$.  Thus, the ordered array
at $f = 1/2$ has three phases: pinned lattice, moving plastic phase,
and moving lattice.

Since a realistic Josephson lattice is certain to have some disorder,
we have also carried out a limited number of calculations for an array
at $f = 1/2$ with weak disorder in the critical currents.
Specifically, we assume that the critical currents are independent
random variables uniformly distributed between $0.9I_c$ and $1.1I_c$.
The resulting phase diagram, for a single realization of disorder, is
shown in Fig.\ \ref{fig:disorder}.  It is calculated using the same
techniques as for ordered arrays.  Once again, we see clear evidence
of three phases: pinned lattice, plastic phase, and moving lattice.
These have characteristics similar to those in the ordered case.  For
example, the moving lattice phase has a finite transverse critical
current $i_{c\perp}$, which goes to zero near the phase boundary.  In
this $10 \times 10$ sample, there is an even larger finger of $B$
phase interposed between $A$ and $C$ than there is in the ordered
array; once again, we assume that this finger disappears in larger
arrays.  For strongly disordered square arrays at several different
field strengths, a phase diagram resembling ours, though without the
interpolated finger, has been found by Dominguez\cite{dominguez}.  We
have also calculated the voltage power spectra for the phases $B$ and
$C$; they resemble those of Fig.\ 4 in that $P(\omega\tau)$ has peaks
at multiples of a fundamental frequency in the moving lattice, while
the spectra in the plastic phase are continuous and only weakly
dependent on frequency.

\section{Simplified Analytical Models}

\label{sec:analyt}

In a further effort to understand the behavior found numerically, we
have considered two simple analytical models.  In this section, we
give a brief description of the models used.  As will be seen, while
each can reproduce {\em some} of the numerical results, neither
generates all the details of the simulations.

\subsection{Four-Plaquette Unit Cell}

Our first analytical model is a slight generalization of an approach
previously used by Rzchowski {\it et al}\cite{rzchowski}.  They
consider the dynamics of a fully frustrated $N \times N$ array of
overdamped resistively-shunted junctions.  To treat this system
analytically, they assume that the dynamical state is simply a
periodic repetition of a square four-plaquette unit cell.  It has long
been known\cite{teitel} that the ground state of the fully frustrated
lattice has such a unit cell, corresponding to the checkerboard vortex
pattern shown in Fig.\ \ref{fig:geom}.  If we maintain this assumption
of periodicity, the equations of Ref.\ \cite{rzchowski} are readily
extended to to the case of currents applied at an angle to the
plaquette edges.  The resulting equations take the form
\begin{eqnarray}
\gamma + \alpha + \beta + \delta = \pi (mod 2\pi) \label{eq:2x2a}\\
-\sin\beta - \frac{d\beta}{dt^\prime} -\sin\delta -
\frac{d\delta}{dt^\prime} +\sin\gamma+\frac{d\gamma}{dt^\prime} + \sin
\alpha + \frac{d\alpha}{dt^\prime} = 0 \label{eq:2x2b}\\
\frac{d\gamma}{dt^\prime} + \sin\gamma -\frac{d\alpha}{dt^\prime} -
\sin\alpha = I_{tot,x} \label{eq:2x2c} \\ \frac{d\beta}{dt^\prime} +
\sin\beta -\frac{d\delta}{dt^\prime} -\sin\delta = I_{tot,y}
\label{eq:2x2d}
\end{eqnarray} 
where $\alpha$, $\beta$, $\delta$, and $\gamma$ are the four
inequivalent gauge-invariant phase differences describing the bonds of
the $2 \times 2$ primitive cell (cf.\ Fig.\ 1 of Ref.\
\cite{rzchowski}), $t^\prime = t/\tau$ is a dimensionless time, and
$I_{tot,x}$ and $I_{tot,y}$ are the total bias currents in the $x$ and
$y$ directions per $2 \times 2$ superlattice cell (in units of the
single-junction critical current).

We have solved these equations numerically, first reducing the system
to three variables and then employing the same integration algorithm
described above. In comparing with the equations for our first set of
simulations, note that the quantity $I_{tot,\alpha} = 2i_{\alpha}
(\alpha=x,y)$ is {\em twice} the current injected into each boundary
grain.  The resulting phase diagram is shown in Fig.\
\ref{fig:2x2phase}.  As in the previous phase diagrams calculated in
this paper, there are regions (denoted $A$, $B$, and $C$) in which
none, one, or both of the time-averaged voltages $v_x$ and $v_y$ are
nonzero.  Despite the reduction in number of variables and the
enforced symmetry of this simulation, the voltages in regime $C$ are
still aperiodic in time, with continuous power spectra at most points
in regime $C$.  A representative power spectrum for $v_x(t^\prime)$ is
shown in the inset to Fig.\ \ref{fig:2x2phase}.  It is calculated for
a point in region $C$ indicated by a triangle.

When the current is applied along the $x$ axis, the critical current
is close to the analytically computed\cite{teitel} value $i_x =
\sqrt{2}-1$.  By contrast, when $i_x >> 1$, the boundaries of region
$B$ asymptotically approach the line $i_y = 1$.  In general, we
conclude that this simplified version of the dynamics has some, but
not all, features of the large array.  In particular it does have a
region of ``plastic flow'' at large $i_x$ and $i_y$.  However, it
fails to reproduce the region of plastic flow interposed between
phases $A$ and $B$ and seen in our larger-scale simulations.

\subsection{Single Vortex in an ``Egg-Carton'' Potential}

Our second analytical model is even simpler.  It refers, not to an
entire lattice of vortices, but to a {\em single} vortex moving in the
``egg-carton'' potential formed by the lattice.  According to Lobb,
Abraham, and Tinkham\cite{lobb}, a single vortex can be viewed, to a
good approximation, as moving in a potential of the form
\begin{equation}
V(x, y) = -V_0\left[\cos\left(\frac{2\pi x}{a}\right) +
\cos\left(\frac{2\pi y}{a}\right)\right],
\end{equation}
where $a$ is the lattice constant of the array and $V_0$ is the depth
of the potential felt by a single vortex.  This potential has minima
at ${\bf r} \equiv (x, y) = (n_1a, n_2a)$ where $n = 0, \pm 1, \pm 2$,
\dots.  [The grains of the Josephson lattice, in these coordinates,
are located at $\left( (n_1+\frac{1}{2})a, (n_2 +
\frac{1}{2})a\right)$, and correspond to {\em maxima} of the vortex
potential.]

The current-voltage characteristics of this model are readily
calculated.  The Magnus force on a vortex due to an external current
density {\bf J} may be written (taking $\hat{z}$ as the direction
perpendicular to the array)
\begin{equation}
{\bf F}_{ext} = \Phi_0\hat{z} \times {\bf J}/c,
\end{equation}
where {\bf J}, in this two-dimensional system, represents a current
per unit {\em length}.  In the steady state, this force has to be
balanced by two other forces: the gradient of the egg-carton potential
energy and the frictional force experienced by the vortex moving
through the lattice.  This condition may be written
\begin{equation}
{\bf F}_{ext} - {\bf \nabla}V({\bf r}) -\eta \dot{{\bf r}} = 0,
\end{equation}  
or explicitly, in component form,
\begin{eqnarray}
\eta \dot{x} = -\frac{\Phi_0 J_y}{c} - \frac{2\pi}{a}V_0\sin\left(\frac{2\pi
x}{a}\right) 
\label{eq:xdot}
\\
\eta\dot{y} = \frac{\Phi_0 J_x}{c} - \frac{2\pi}{a} V_0\sin\left(\frac{2\pi
y}{a}\right).
\label{eq:ydot}
\end{eqnarray}

Given the vortex velocities, the electric fields may be written down
by using the relation between vortex velocity and electric field.
Specifically, the voltage drop between any two points $P_1$ and $P_2$ is
given by
\begin{equation}
\Delta V_{12} = 2\pi n v_{\perp} L\frac{\hbar}{2e},
\end{equation}
where $v_{\perp}$ is the component of vortex velocity perpendicular to
the line joining $P_1$ and $P_2$, $L$ is the distance between $P_1$ and $P_2$,
and $n$ is the vortex number density per unit area.  Here we have used
the fact that the phase difference between $P_1$ and $P_2$ changes by
$2\pi$ every time a vortex crosses that line, and have also availed
ourselves of the Josephson relation between voltage and phase.  We
therefore deduce the following expressions for the components of
electric field:
\begin{eqnarray}
E_x = \frac{hn\dot{y}}{2e} \label{eq:ex} \\
E_y = -\frac{hn\dot{x}}{2e}.
\label{eq:ey}
\end{eqnarray} 

Eqs.\ (\ref{eq:xdot}) and (\ref{eq:ydot}) are identical in form to the
equations for single Josephson junctions, with $x/a$ and $y/a$ playing
the role of the Josephson phase, $2\pi V_0/a$ the role of the critical
current, and $-\Phi_0 J_y/c$ and $\Phi_0 J_x/c$ the roles of the
driving currents.  In view of eqs.\ (\ref{eq:xdot}) and
(\ref{eq:ydot}), we deduce that the time-averaged voltage drop in the
$i$ direction ($i = x, y$) becomes nonzero when $|\Phi_0 J_i/c| > 2\pi
V_0/a$.  As has been shown by ref.\ \cite{lobb}, $V_0$ is related to
the critical current $I_c$ of an individual Josephson junction by
\begin{equation}
V_0 \sim 0.22 \frac{\hbar I_c}{2e}.
\end{equation}
Collecting all this information, we obtain the simple phase diagram
shown in Fig.\ \ref{fig:eggphase} for the voltage drops arising from
motion of a single vortex in a square array.  Once again, there are
three regimes, denoted $A$, $B$, and $C$, where none, one, or both of
the time-averaged voltage drops $\langle v_x\rangle_t$ and $\langle
v_y \rangle_t$ are nonzero.  However, the phase diagram is simplified
by the fact the $\langle v_x\rangle_t$ and $\langle v_y\rangle_t$ are
independent of one another, depending only on $i_x$ and $i_y$
respectively.  The phase boundaries correspond to the vertical and
horizontal lines $v_x \sim 0.11$ and $v_y \sim 0.11$.

The results of Fig.\ \ref{fig:eggphase} are obviously oversimplified
compared to the numerical diagrams of Fig.\ 2.  In addition to the
other obvious differences, the voltages in region $C$ are not chaotic,
but are instead just the superposition of two independent voltages in
the $x$ and $y$ directions, each of which has its own fundamental
frequency and harmonics of that fundamental.  The two fundamentals
may, of course, be incommensurate depending on the values of $\langle
v_x\rangle_t$ and $\langle v_y\rangle_t$.

\section{Discussion and Conclusions}

\label{sec:discuss}

The present calculations clearly show that a fully frustrated array of
overdamped Josephson junctions exhibits at least three dynamical phases as
a function of the two orthogonal driving currents $i_x$ and $i_y$: a pinned
vortex lattice, a plastic phase characterized by a continuous power spectrum
for both $v_x$ and $v_y$, and a moving vortex lattice with only one of
the voltages $v_x$ and $v_y$ nonzero.  In this last phase, the power spectrum,
at least for the limited lattice sizes we have investigated, contains only
harmonic multiples of a fundamental frequency.  Weak disorder in the critical
currents appears not to change this phase diagram greatly.

Regarding our phase diagram, it is natural to ask whether the
boundaries between the different phases are analogous to first-order
phase transitions.  Although we have no conclusive evidence, our
numerical results suggest that they may indeed be first-order, rather
than continuous, at least for the ordered lattices.  In support of
this conjecture, we note the occasional occurrence of hysteresis in
our simulations, and of discontinuous jumps between one phase and
another near the phase boundaries.  There is also little evidence that
any quantities, such as the strength of the voltage noise, diverge
near the phase boundaries, as might be expected of a continuous phase
transition.  In this respect, these transitions differ somewhat from
those seen in strongly disordered lattices.

For these ordered arrays in which the vortex lattice is commensurate
with the underlying Josephson array, one might have expected other
types of commensurate-incommensurate transitions as the
angle between the applied current and array symmetry axis is varied.
Such ``magic angle'' effects, with a multitude of commensurate and
disordered phases, are seen in other models in which vortex lattices
are driven through periodic pinning arrays\cite{nori,nori1}.  In the
present case, we have seen no clear evidence of any phases other than
the three shown in our phase diagram.  Possibly, such additional
phases would appear if we studied larger arrays in which such delicate
effects would be more stable.  On the other hand, our
open boundary conditions may constitute such a strong perturbation on the
periodic array that such commensurability effects would be suppressed
even for very large lattices.

Finally, we comment briefly on the power spectra found in our
simulations.  In the moving lattice phase $B$, the voltage power
spectra appear to contain only multiples of a fundamental frequency.
Such a power spectrum represents an array which is phase-locked, and
hence would radiate power only at multiples of that fundamental
frequency.  In this case, phase-locking clearly occurs {\em between}
rows as well as {\em along} rows of junctions parallel to the voltage.
Such phase locking must be present because it is required in order to
produce the observed relation between $\langle v_x\rangle_t$ and the
fundamental frequency.  Furthermore, such locking survives weak
disorder in the critical currents.  The effectiveness of a magnetic
field in producing phase locking in square arrays has been discussed
previously\cite{fil95,trees}.  The present work provides additional
evidence that a field $f = 1/2$ is effective in producing phase
locking, even when the applied current has nonzero components parallel
to both array axes\cite{aranson}.

In conclusion, we have numerically investigated the dynamic phases of
fully-frustrated square Josephson junction arrays driven by two
independent, orthogonal currents. We find three phases: stationary
lattice, driven lattice, and plastic flow. These phases appear also in
weakly disordered arrays. We find that our numerical results
can be partly understood by two
simple analytical models, which show some of the features of the full
simulation.

\section{Acknowledgments}

This work has been supported by the National Science Foundation, Grant
No. DMR97-31511, and by the Midwest Superconductivity Consortium at Purdue
University through Grant DE-FG 02-90 ER45427.  We thank Profs.
Predrag Cvitanovic and Franco Nori for useful conversations.

\newpage

\begin{figure}
\caption{Schematic of the geometry for calculating IV characteristics.
Currents $i_x = I_x/I_c$ and $i_y=I_y/I_c$ are injected into each
grain (shaded circles) on left-hand and lower edges of the square
array, and extracted from right-hand and upper edges, as shown.
Time-dependent but spatially averaged voltage drops $v_x(t^\prime)$
and $v_y(t^\prime)$are calculated across the array in the $x$ and $y$
directions.  Crosses denote vortex locations, as defined in the text,
for the ground-state checkerboard pattern at $f= 1/2$.}
\label{fig:geom}
\end{figure}

\begin{figure}
\caption{Calculated zero-temperature ``dynamical phase diagram'' for
(a) $10 \times 10$ and (b) $20 \times 20$ square Josephson arrays with
currents applied in both the $x$ and $y$ directions.  Region $A$:
pinned vortex lattice ($\langle v_x\rangle_t = \langle v_y\rangle_t =
0$).  Region $B$: moving vortex lattice phase ($\langle v_x \rangle_t
= 0, \langle v_y\rangle_t > 0$, or $\langle v_y\rangle_t = 0, \langle
v_x\rangle_t > 0$).  Region $C$: moving plastic vortex phase ($\langle
v_x\rangle_t > 0, \langle v_y\rangle_t > 0$).  The phase boundaries
were determined numerically at the calculated points, as described in
the text; the smooth lines are interpolations between these points.
The phase diagram is assumed symmetric about a 45$^o$ line; diamonds
denote points which were calculated for $i_x > i_y$; circles, for $i_y
> i_x$.  The sliver of phase $C$ between $A$ and $B$ in (a) is a
finite-size artifact, as shown in (b) where it is seen to be absent in
a $20 \times 20$ array. Open triangles in (b) indicate currents
($i_x$, $i_y$) at which the voltage traces and power spectra of
Figs. \ref{fig:Vtrace} and \ref{fig:Vpow} were calculated.}
\label{fig:phased}
\end{figure}    

\begin{figure}
\caption{Time-dependent voltage traces $v_x(t^\prime)$ and
$v_y(t^\prime)$ (upper and lower parts of each panel) for several
points in the phase diagram of a $20 \times 20$ array at $f = 1/2$.
(a) [$i_x=0.351, i_y=0.079$] and (b) [$i_x=405, i_y=0.294$], represent
points in the plastic phase $C$; (c) [$i_x=0.508, i_y=0.210$] and (d)
[$i_x=0.469, i_y=0.287$] are points in the moving-lattice phase $B$.  As
described in the text, $v_x(t^\prime)$ represents the difference
between the average voltages on the left-hand and right-hand edge of
the array; $v_y(t^\prime)$ is the difference between the average
voltages on the top and bottom edges of the array.  $t^\prime =
t/\tau$ is a dimensionless time; $\tau$ is a natural time unit defined
in the text.}
\label{fig:Vtrace}
\end{figure}

\begin{figure}
\caption{Calculated voltage power spectra $P(\omega\tau)$ for the
voltages $v_x(t^\prime)$ and $v_y(t^\prime)$ of Fig.\ \ref{fig:Vtrace}
(a) - (d). Here $\omega$ is the angular frequency. In each panel, the
power spectrum for $v_x$ lies above that for $v_y$.}
\label{fig:Vpow}
\end{figure}

\begin{figure}
\caption{Calculated phase diagram for a $10 \times 10$ array at $f =
1/2$ with one realization of weak disorder.  Rather than all being
equal, the critical currents are chosen at random from a distribution
which is uniform in the interval (0.9$I_c$, 1.1$I_c$).}
\label{fig:disorder}
\end{figure}

\begin{figure}
\caption{Calculated phase diagram for a square array at $f = 1/2$,
obtained by solving the overdamped dynamical equations for a $2 \times
2$ unit cell which is assumed to be repeated throughout the lattice
[eqs. \ref{eq:2x2a}-\ref{eq:2x2d}].  The phase diagram is
shown as a function of the driving currents $i_x$ and $i_y$ in the $x$
and $y$ directions.  As in previous phase diagrams, there are regions
where none, one, or both of the two time-averaged voltages $\langle
v_x\rangle_t$ and $\langle v_y\rangle_t$ are nonzero. Inset shows
power spectrum of $v_x(t^\prime)$ for the point indicated by the open
triangle ($i_x=1.348, i_y=1.115$).}
\label{fig:2x2phase}
\end{figure}

\begin{figure}
\caption{Calculated phase diagram for a single vortex moving in the
egg-carton potential produced by a square array of overdamped
Josephson junctions.  The notation is the same as in Fig.\
\ref{fig:2x2phase}.  Once again, there are regions where none, one, or
both of the time averaged voltages $\langle v_x \rangle_t$ and 
$\langle v_y\rangle_t$ are nonzero.}
\label{fig:eggphase}
\end{figure}

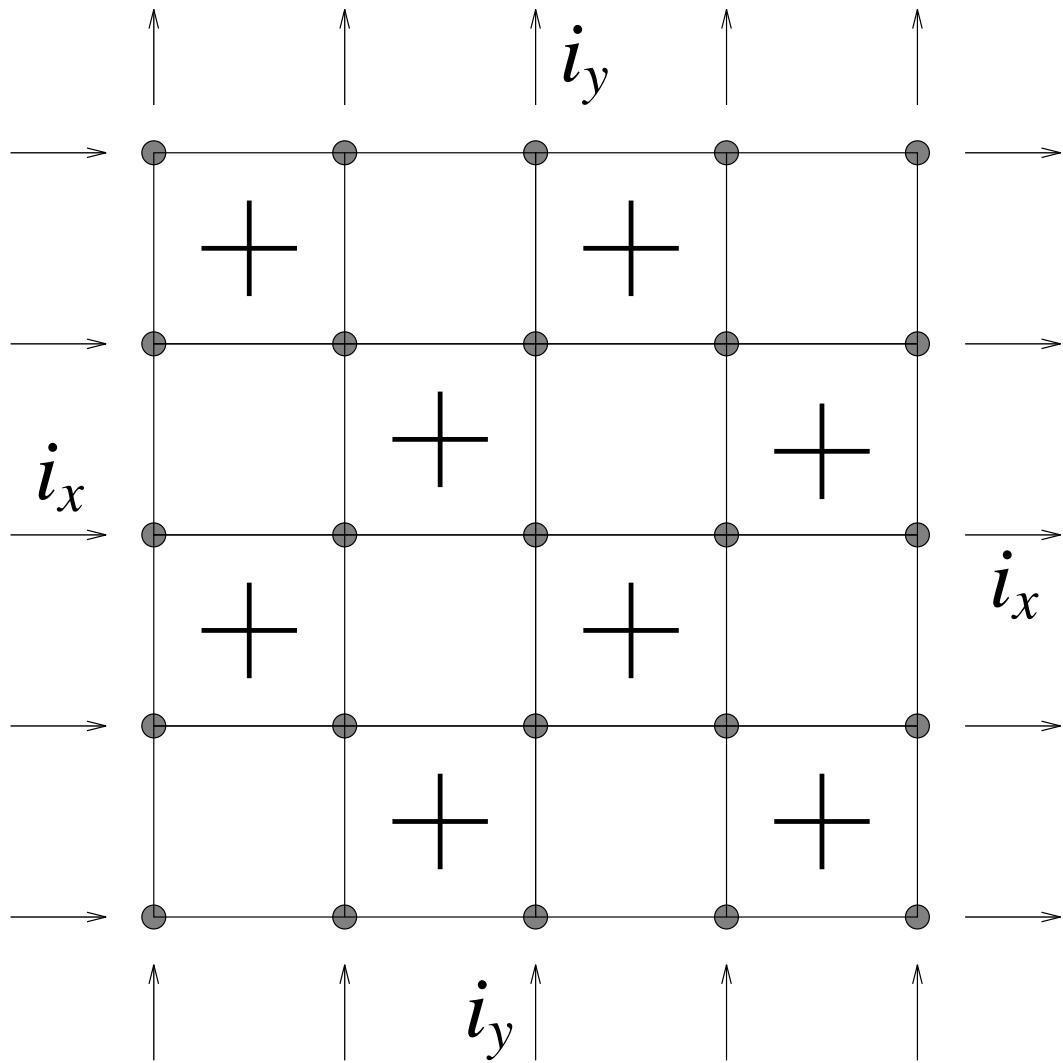
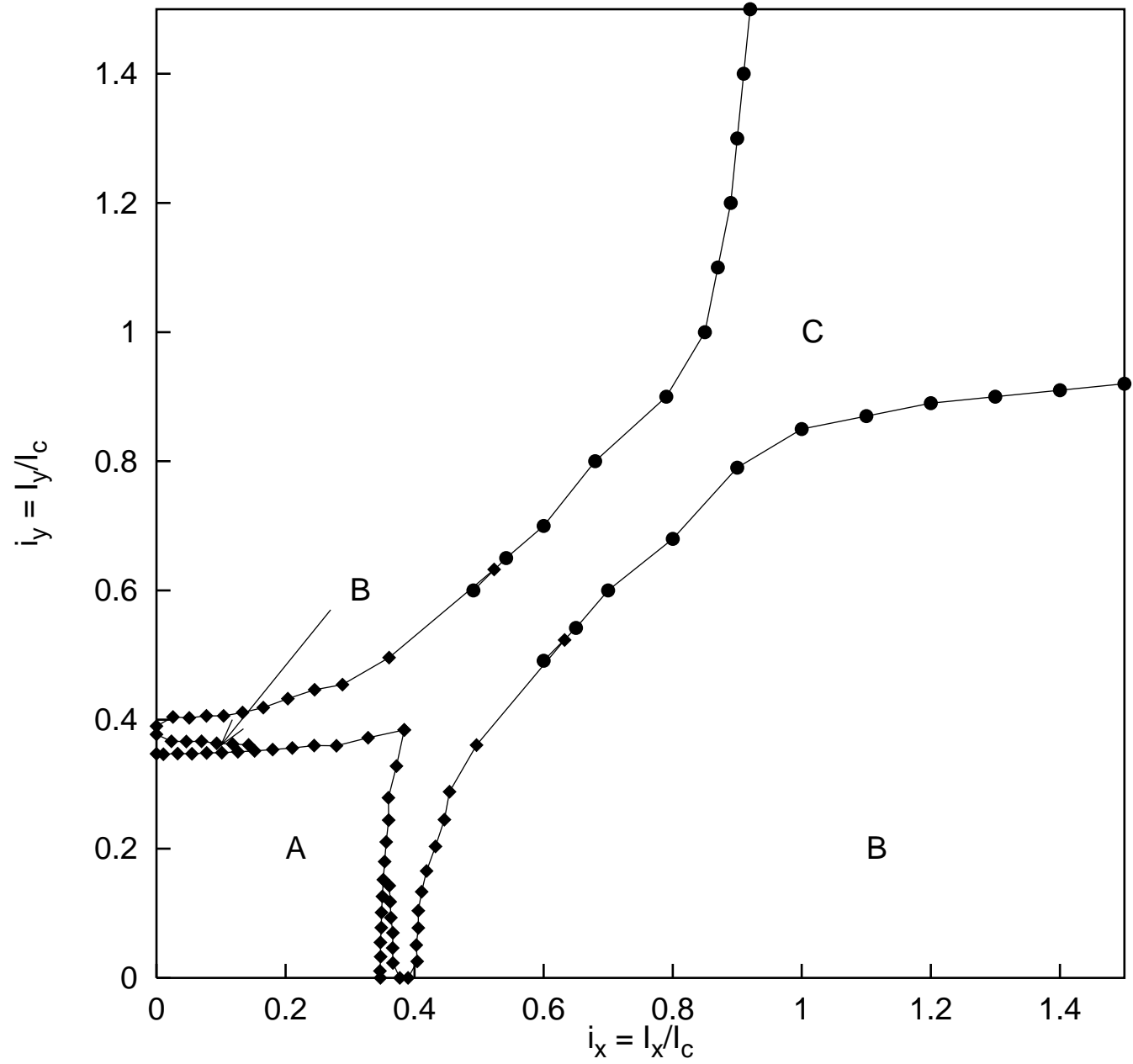
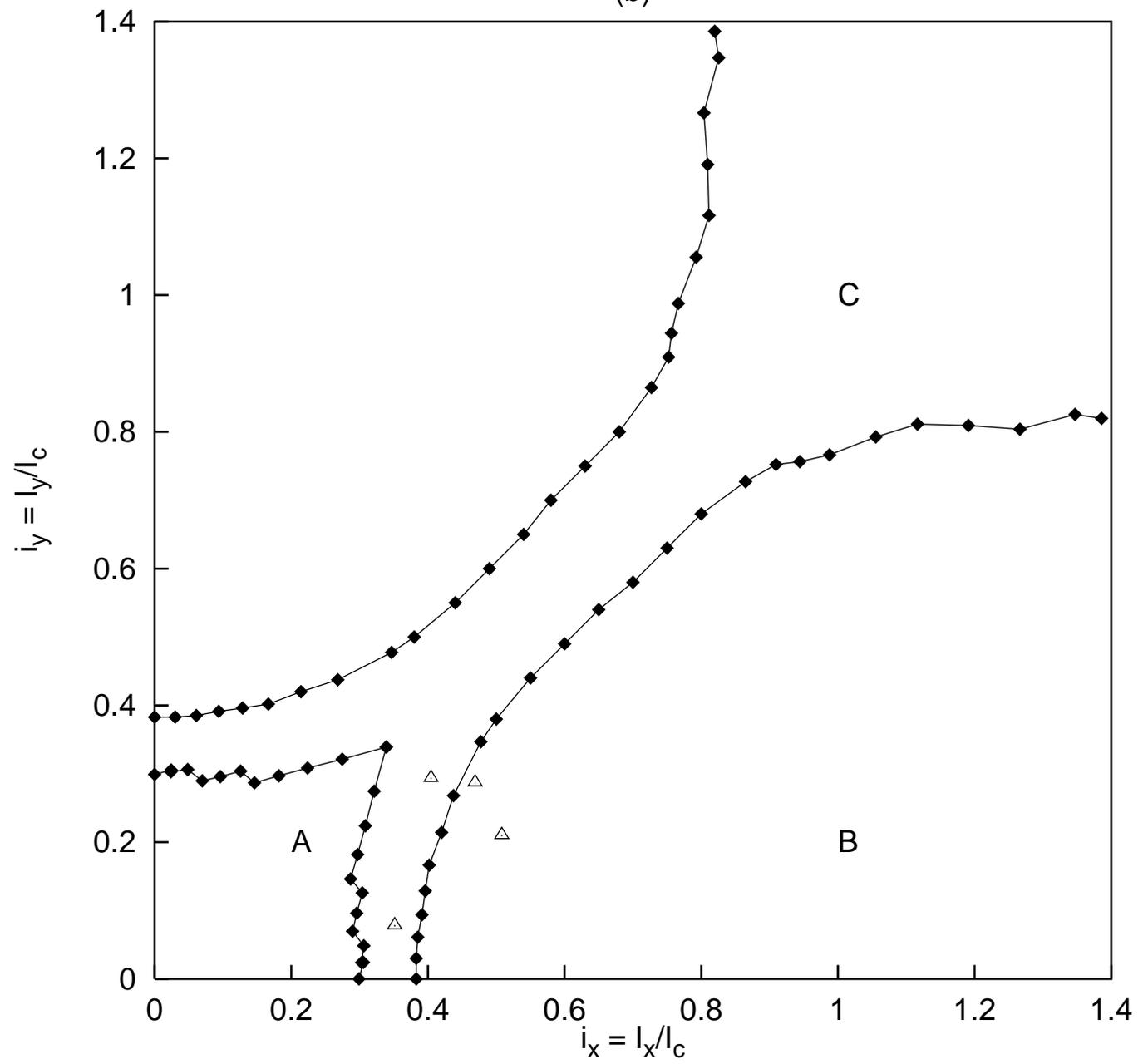
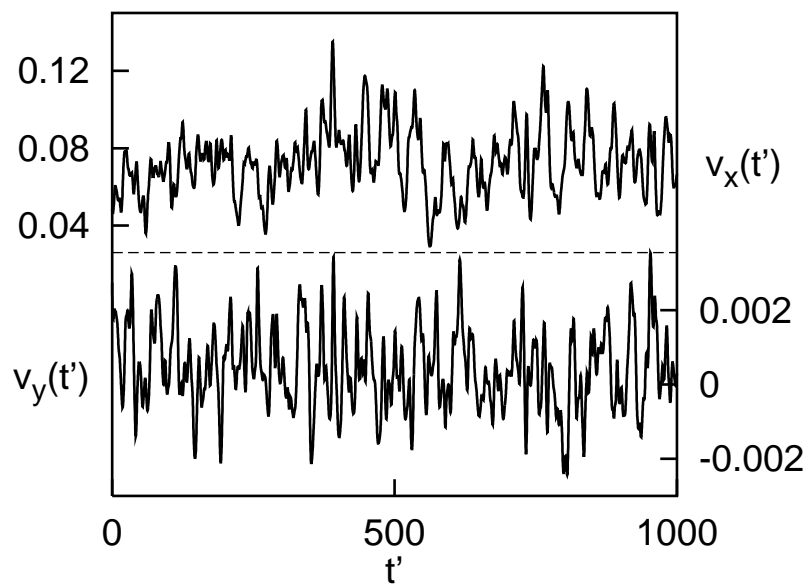
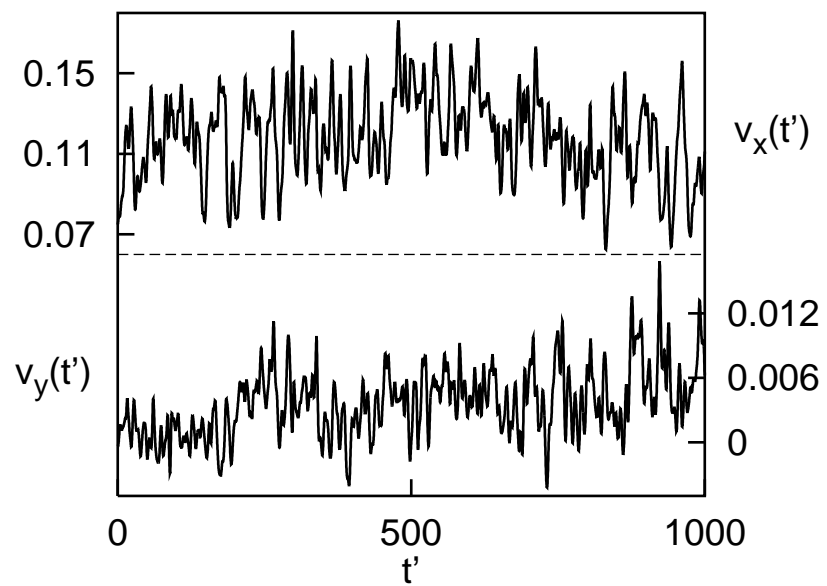
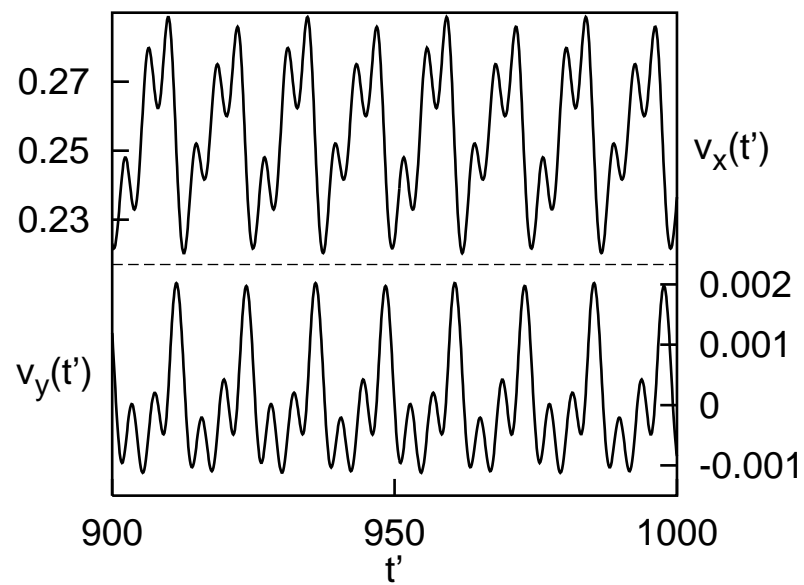
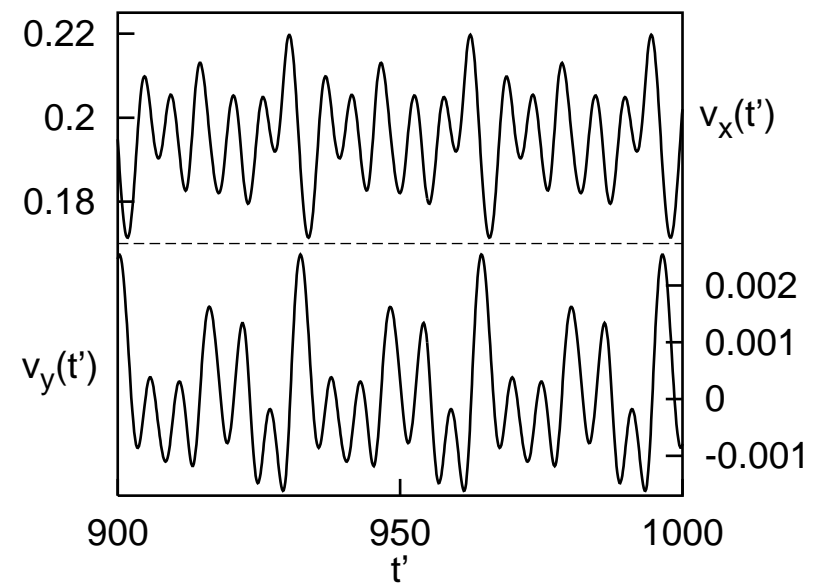
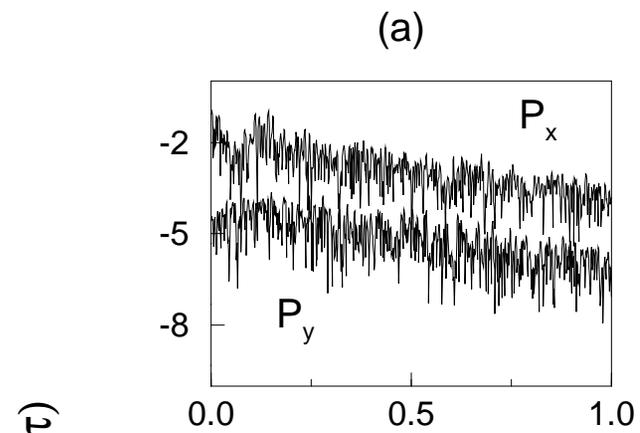
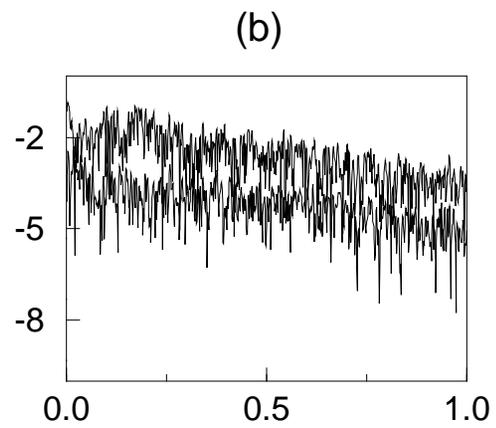
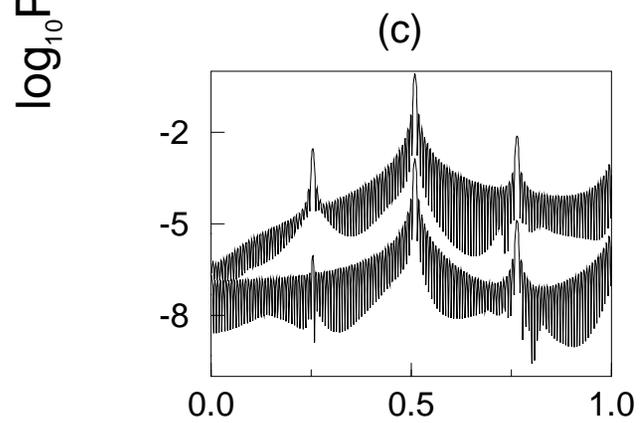
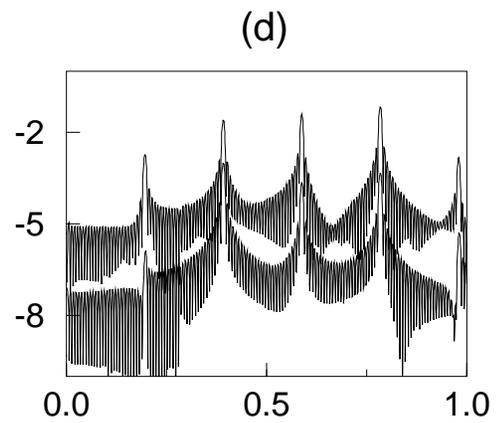
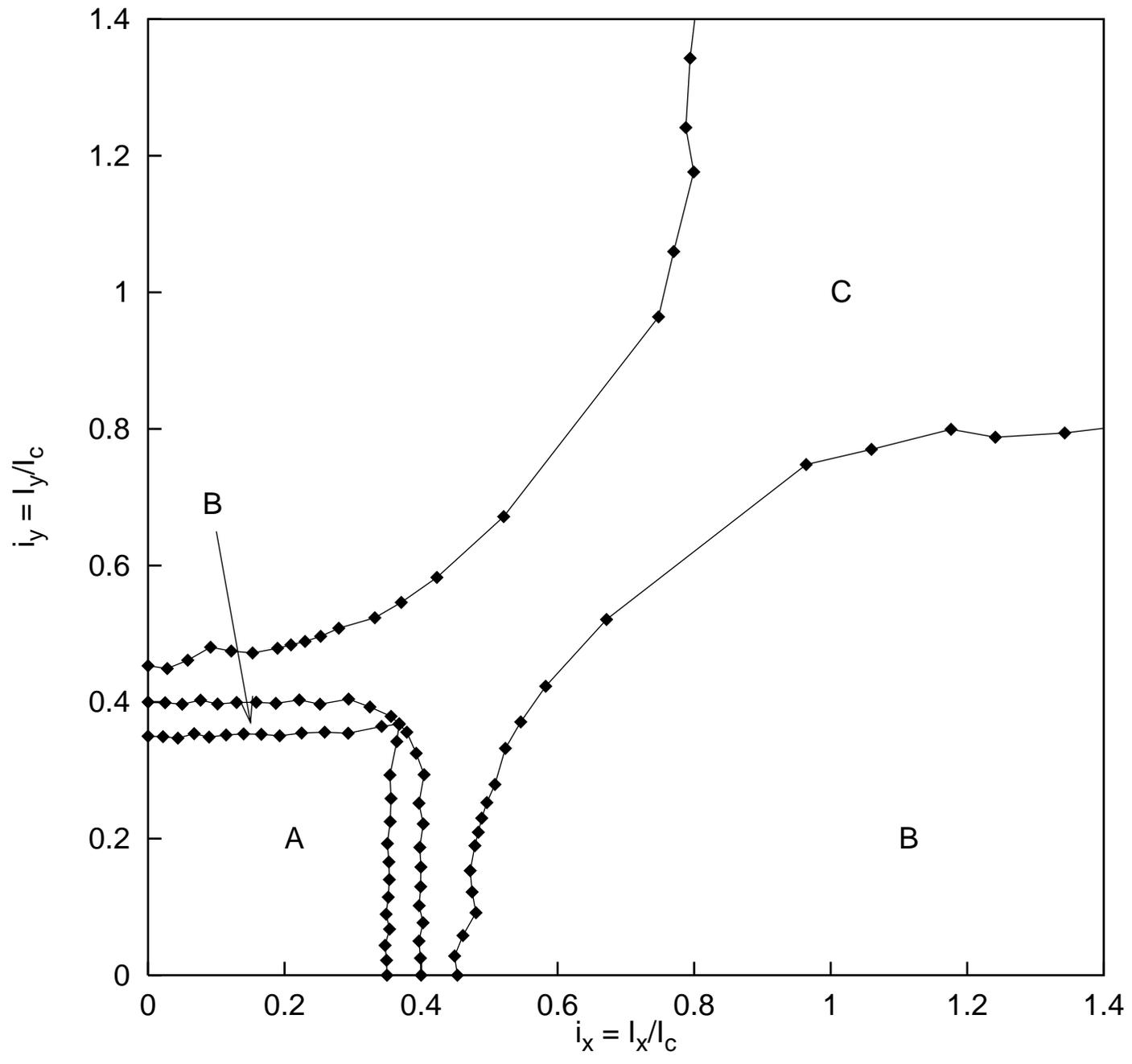
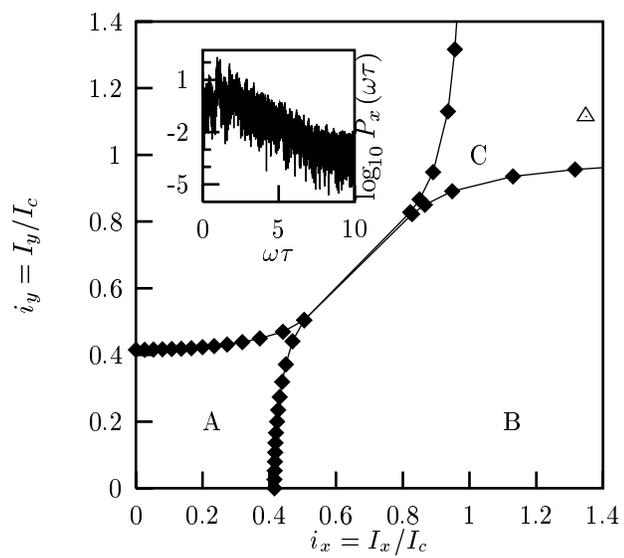
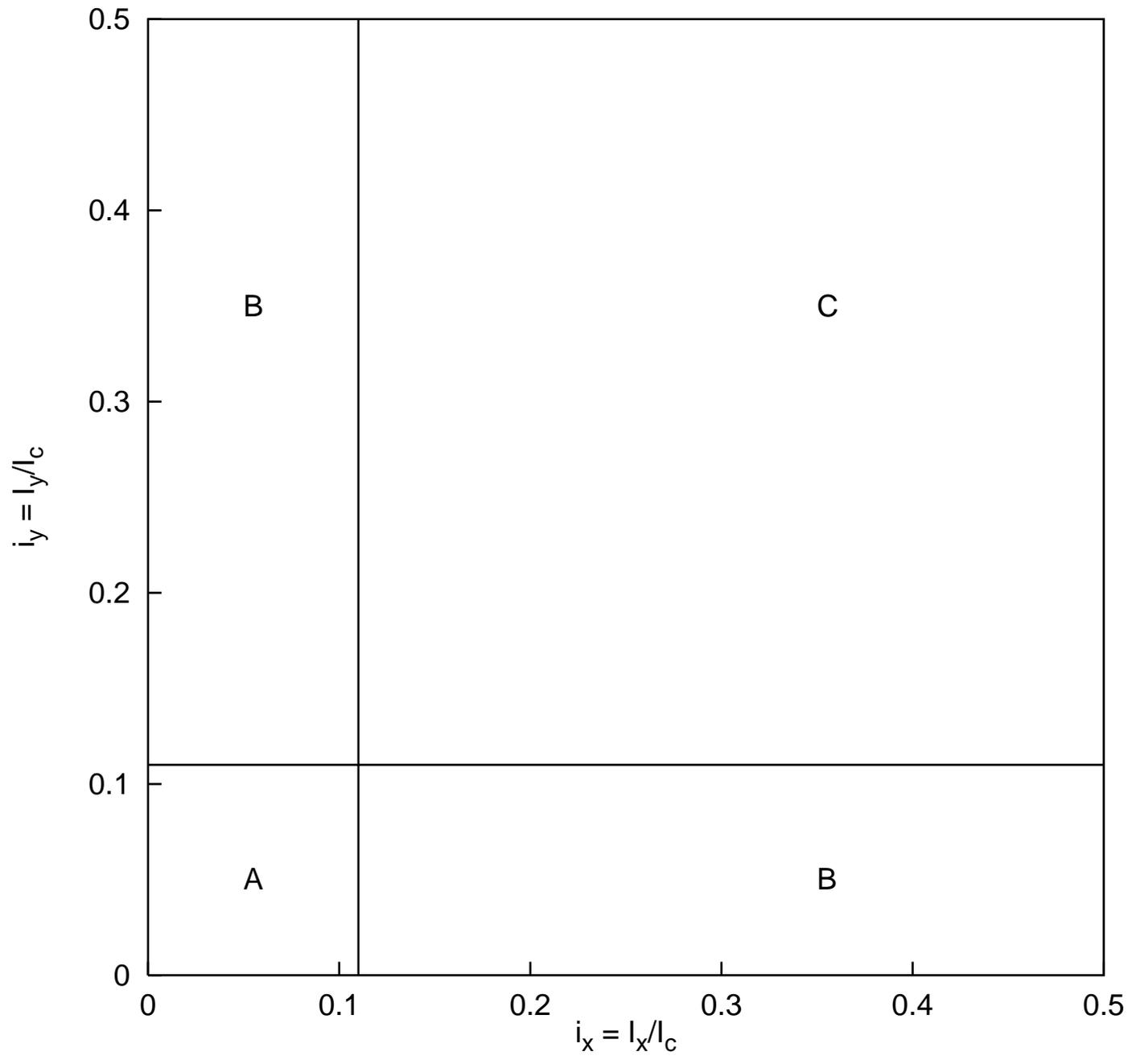
\end{document}